\documentclass[vecphys]{svmult}

\usepackage{makeidx}         
\usepackage{graphicx}        
\usepackage{multicol}        
\usepackage[bottom]{footmisc}
\makeindex             

\begin{document}

\title*{States and transitions in black-hole binaries}
\author{Tomaso M. Belloni}
\institute{INAF - Osservatorio Astronomico di Brera, Via E. Bianchi 46, I-23807 Merate, Italy
\texttt{tomaso.belloni@brera.inaf.it}}

\maketitle

\abstract{With the availability of the large database of black-hole transients from the
Rossi X-Ray Timing Explorer, the observed phenomenology has become very complex. The original classification of the properties of these systems in a series of static states sorted by mass accretion rate proved not to be able to encompass the new picture. I outline here a summary of the current situation and show that a coherent picture emerges when simple properties such as X-ray spectral hardness and fractional variability are considered. In particular, fast transition in the properties of the fast time variability appear to be crucial to describe the evolution of black-hole transients.
Based on this picture, I present a state-classification which takes into account the observed transitions. I show that, in addition to transients systems, other  black-hole binaries and Active Galactic Nuclei can be interpreted within this framework. The association between these states and the physics of the accretion flow around black holes will be possible only through modeling of the full time evolution of galactic transient systems.
}

\section{Introduction}
\label{sec:1}

The presence of two different ``states" in the X-ray emission of the first black-hole candidate, Cygnus~X-1, was realized in the early 70Õs with the Uhuru satellite \cite{Tanan72}. A transition lasting less than one month was observed from a soft spectrum to a hard spectrum: the source intensity decreased by a factor of 4 in the 2-6 keV band and increased by a factor of 2 in the 10-20 keV band. The associated radio source was not detected before the transition and appeared at a flux at least 3 times higher after the transition. Because of the large flux swing below 10 keV, where most of the instrument response was, this led to the identification of a ``high" state (with soft spectrum and no radio detection) and a ``low" state (with hard spectrum and associated radio source). In its ``low" state, Cyg~X-1 was observed to show strong aperiodic noise \cite{Terrell72,Nolan81}, while the soft state is characterized by a much reduced noise level.

As we know now, most Black-Hole Binaries (BHB) are transient sources (Black-Hole Transients, BHT), whose detection depends strongly on the availability of all-sky monitors and wide-field instruments. The first BHT was A~0620-00\index{A~0620-00}, discovered with the Ariel satellite in 1975 \cite{Elvis75}. As its flux reached 50 Crab at peak, the spectral distribution could be followed throughout the outburst \cite{Ricketts75}. The spectrum was hard at the start of the outburst, while the flux peak was observed to be caused by a strong soft ($<$10 keV) enhancement, while the hard flux dropped. The two states recognized in Cyg X-1 appeared to be present also here, as recognized by \cite{Coe76}.

Other sources appeared to follow this bi-modal state classification. The two persistent systems LMC X-1 and LMC X-3 were always observed in the high state (although it is now recognized that LMC X-3 shows occasional transitions to a harder state (see \cite{Wilms01,Gotz06}). The bright source GX 339-4 was not detected with Uhuru and was discovered with OSO-7 \cite{Markert73}. The early observations showed that the source was extremely variable over time scales of 100 days, alternating between three states: a ``high" state with a soft spectrum, a harder ``low" state and an ``off" state, which was later recognized as a low-flux extension of the low state (see \cite{Ilovaisky86}). The similarity of GX~339-4 and Cyg~X-1 was also strengthened by the observation of similar strong aperiodic variability \cite{Samimi79}. Since the names ``low'' and ``high'' derive from the 1-10 keV flux and the fluxes at higher energies reverse, I will refer to them as ``low-hard'' (LHS) and ``high-soft'' (HSS) respectively (see Chap. 2).

The all-sky monitor on board the Ginga satellite allowed the discovery of new X-ray transients, which could then be followed-up with extensive observations with its large proportional counters. Three major bright transients were observed, GS~1124-684, GS~2000+251 and GS~2023+338 (see \cite{TanLew95}). The latter showed very unusual properties and dramatic variability caused by variable intrinsic absorption. GS~2000+251 showed low and high states similar to those already known. More interesting was the case of GS~1124-684\index{GS~1124-684}. Here, in addition to the two known states, an additional state with different properties was found \cite{Kitamoto92}. This new state had already been recognized in earlier observations of GX~339-4 \cite{Miyamoto91}: since it appeared at the brightest flux levels, it was dubbed ``very-high" state (VHS). Its spectral properties were a mixture of those observed in the high and low states, while the timing properties appeared very complex. Low-frequency Quasi-Periodic Oscillations (QPO) were observed, but not at all times, and fast transitions could be seen (see \cite{Miyamoto94,Takizawa97}). Properties similar to those of the ``very-high" state were seen in the same two sources, but during observations at a much lower luminosity \cite{Belloni97,Mendez97}: a new ``intermediate'' state (IMS) was therefore proposed.

Overall, the amount of information available until 1995 was scarce and it was difficult to derive a coherent picture (see \cite{TanLew95} for a review).
At the very end of 1995, the Rossi X-Ray Timing Explorer was launched \cite{RXTE}. The presence of an all-sky monitor (ASM), a large proportional counter (PCA) and a high-energy instrument (HEXTE), together with an extreme flexibility of operation, make it an ideal mission for the study of bright black-hole binaries
\footnote{Notice that the original denomination of ``black-hole candidate" (BHC) has at some point been replaced with ``black-hole binary'' (BHB) without any qualitative breakthrough to justify the change in nomenclature.}, most of which are of transient nature. A vast database is available, which of course is enhanced by the presence of (sparser) observations by other X-ray missions. The large amount of new information has naturally resulted in a burst of publications. Since the observed phenomenology is complex, it resulted in a number of different classifications in terms of source states, often evolving with time and difficult to compare with each other.  In the following, I will present the current situation, with the aim of guiding the reader through the jungle of source properties and states. My approach will be touching the basic properties, intentionally ignoring all inevitable complications, in order to give an overview of the general picture that has emerged, pointing to selected publications where the subject can be examined in more detail. I concentrate on the X-ray properties, since the connection with other energy bands and with the jet ejection is discussed in other chapters. The use of the term BHB instead of microquasar is intentional and aimed at denote the complete class of sources, whether relativistic jet ejections have been observed or not. The ubiquitous presence of radio emission, at least in some states, seems to indicate that the two definitions probably point to the same set of objects.

\section{The fundamental diagrams\index{Hardness-Intensity Diagram (HID)}\index{Hardness-rms diagram (HRD)}}
\label{sec:2}
While the HSS and the LSS are relatively well defined and identified (see Chapter 2), everything else is still a matter of discussion. The light curves of Black-Hole Transients are quite varied and even the same source can exhibit very different time evolutions of their flux during different outbursts (see \cite{HomBel05,McRembook}). However, the spectral hardness (defined as the ratio of observed counts in two energy bands) and the integrated fractional rms variability are quantities which exhibit considerable regularity (\cite{Belloni05,HomBel05}) and have the additional advantage of being model-independent. Since RXTE provides a large database obtained with the same high-area instrument (the PCA), after correction for time variations of the instrument gain, it is possible to compare a large number of sources, comparison which would be impossible combining different instruments. Clearly, the interpretation of hardness is not obvious and need to be supported by spectral fitting, also considering that the observed spectrum consists of the superposition of different components. However, a careful selection of the energy bands allows to separate the major components and avoid complex degeneracies.

\begin{figure}
\centering
\includegraphics[height=9cm]{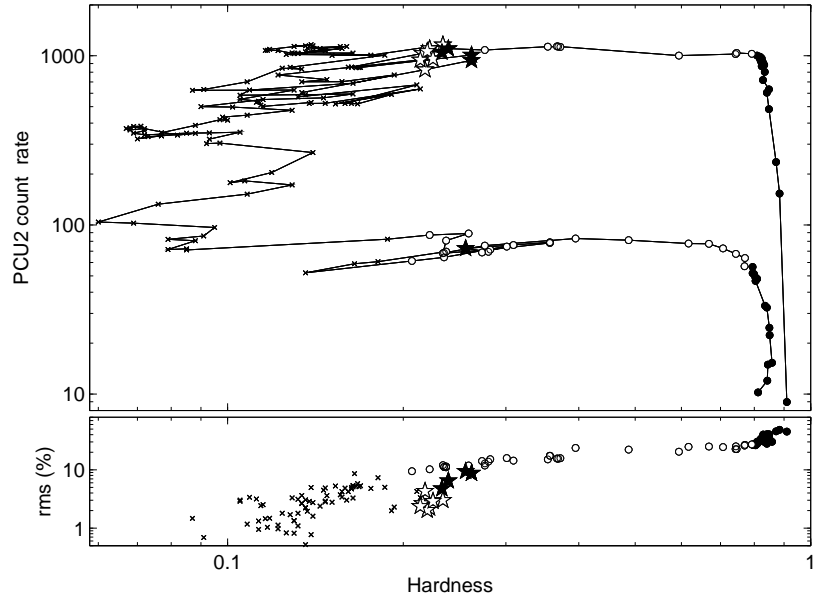}
\caption{
Hardness-Intensity diagram (HID: top panel) and Hardness-Rms Diagram (HRD: bottom panel) for the 2002/2003 outburst of GX~339-4 (adapted from \cite{Belloni05}). PCA count rate is in the energy range 3.8--21.2 keV. Hardness is defined as the ratio of counts in the energy bands 6.3--10.5 and 3.8--6.3 keV. Fractional percentage rms corresponds to the frequency range 0.1--64 Hz and to the full energy range. Symbols correspond to different types of Power Density Spectra (see Sect. \ref{sec:3}): type 1 (filled circles), type 2 (empty circles), type 3 (filled stars), type 4 (empty stars), type 5 (dots)
}
\label{fig:hid339}
\end{figure}

Two diagrams are particularly useful for characterizing the behavior of BHT: the hardness-intensity diagram  (HID), where the total count rate is plotted as a function of hardness, and the hardness-rms diagram (HRD), where the fractional rms, integrated over a broad range of frequencies, is plotted versus hardness. The HID is equivalent to the  color-magnitude diagram of optical astronomy, with the  major difference that for BHT we can follow the movement of a single source on short time scales. A number of diagrams can be produced from parameters obtained from X-ray spectral fitting: these can be considered the equivalent of the Herzsprung-Russell diagram (see e.g. \cite{GDone04,RemMcARAA}). However, for stellar astronomy the emission model is better understood, justifying the loss of model-independence, while theoretical assumptions in the case of BHT are usually far from being agreed upon.

Work on the complete RXTE/PCA sample of BHT shows that, although the time evolution of the parameters can be quite different, these diagrams allow the identification of a surprising number of common properties \cite{HomBel05,HomBel08}.
A full comparative analysis of the BHT in the RXTE archive is presented by \cite{HomBel08}. Here I will base my description on the 2002--2003 outburst of GX~339-4\index{GX~339-4}, as published in two earlier papers \cite{Nespoli03,Belloni05}. This source is clearly the clearest example, as confirmed by the fact that its HID and HRD are extremely similar also for all outbursts observed with RXTE, but the general statements that can be drawn from it apply to all other transients observed by RXTE \cite{HomBel08}.

The HID and HRD of GX~339-4 are shown in Fig. \ref{fig:hid339}. From the q-shaped HID (also referred to as ``turtle head''\index{turtle head}), four distinct branches can be identified, corresponding to the four sides of the `q'. These could be associated to the original four pre-RXTE states: a LHS on the hard branch to the right, a HSS on the soft branch to the left, a VHS on the branch the top and a IMS on the bottom, the last two with large hardness variations. From a complete analysis of the 1998 outburst of XTE~J1550-564, it was shown that the properties of VHS and IMS are very similar in all respects and can take place at different luminosity levels, leading to a unification of these into one likely physical state (\cite{Homan1550}). This leaves us with three separate states. 
However, when timing properties are taken into account, the HRD (bottom panel on Fig. \ref{fig:hid339}) gives a different picture: here most of the points follow  a single correlation: the LHS points have a high level of variability (more than 20\%), the HSS points vary much less ($<$10\%) and the intermediate points are, of course, intermediate. Interestingly, there is a narrow band in hardness where many points have a lower variability than those on the main correlation (shown as stars in Fig. \ref{fig:hid339}). This diagram suggests the presence of two separate states only, corresponding to the points on the main correlation (spanning a large range in hardness) and to those deviating from it. 

\begin{figure}
\centering
   \begin{tabular}{c}
		\includegraphics[height=5.7cm]{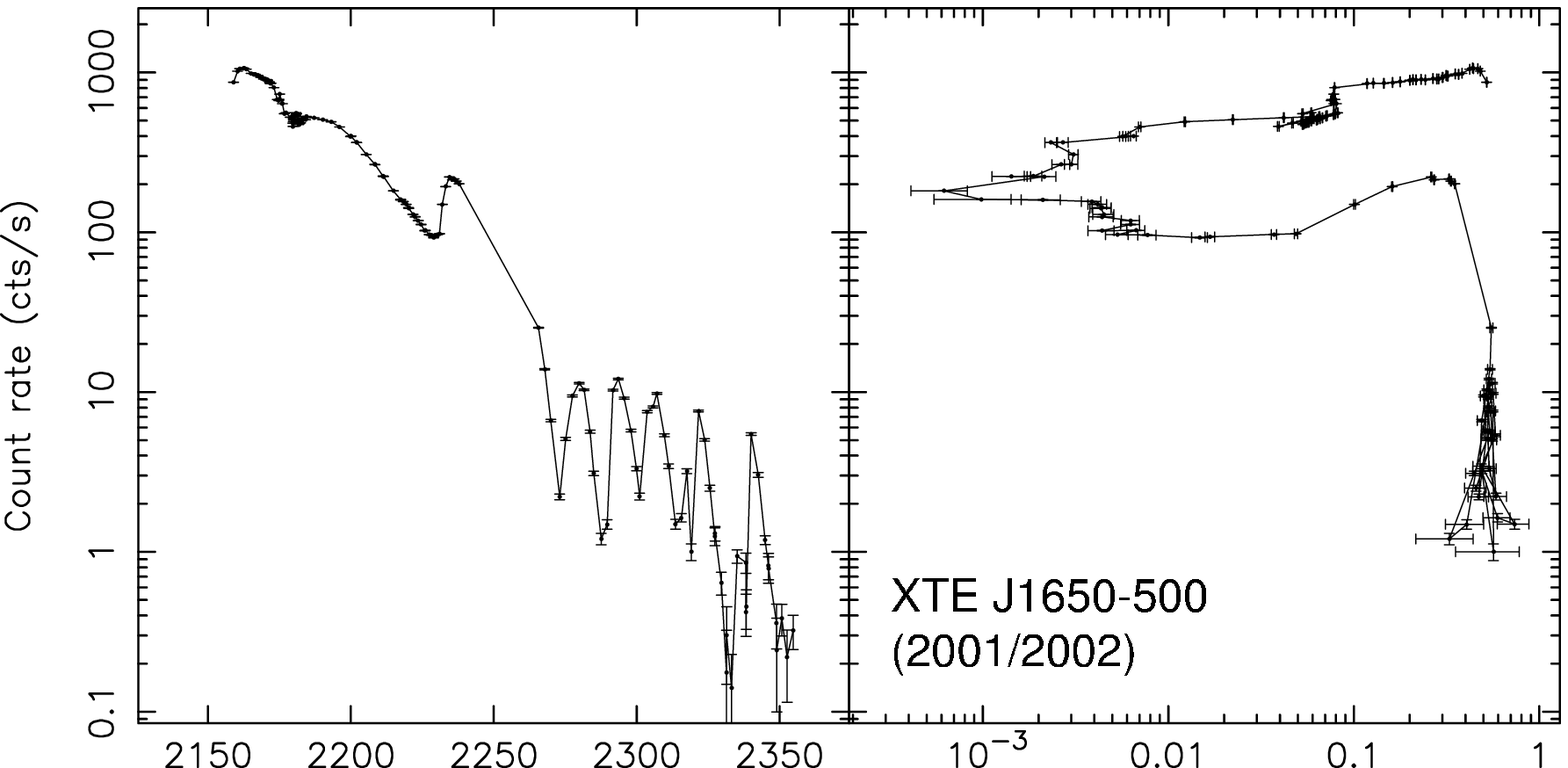}\\
		\includegraphics[height=5.82cm]{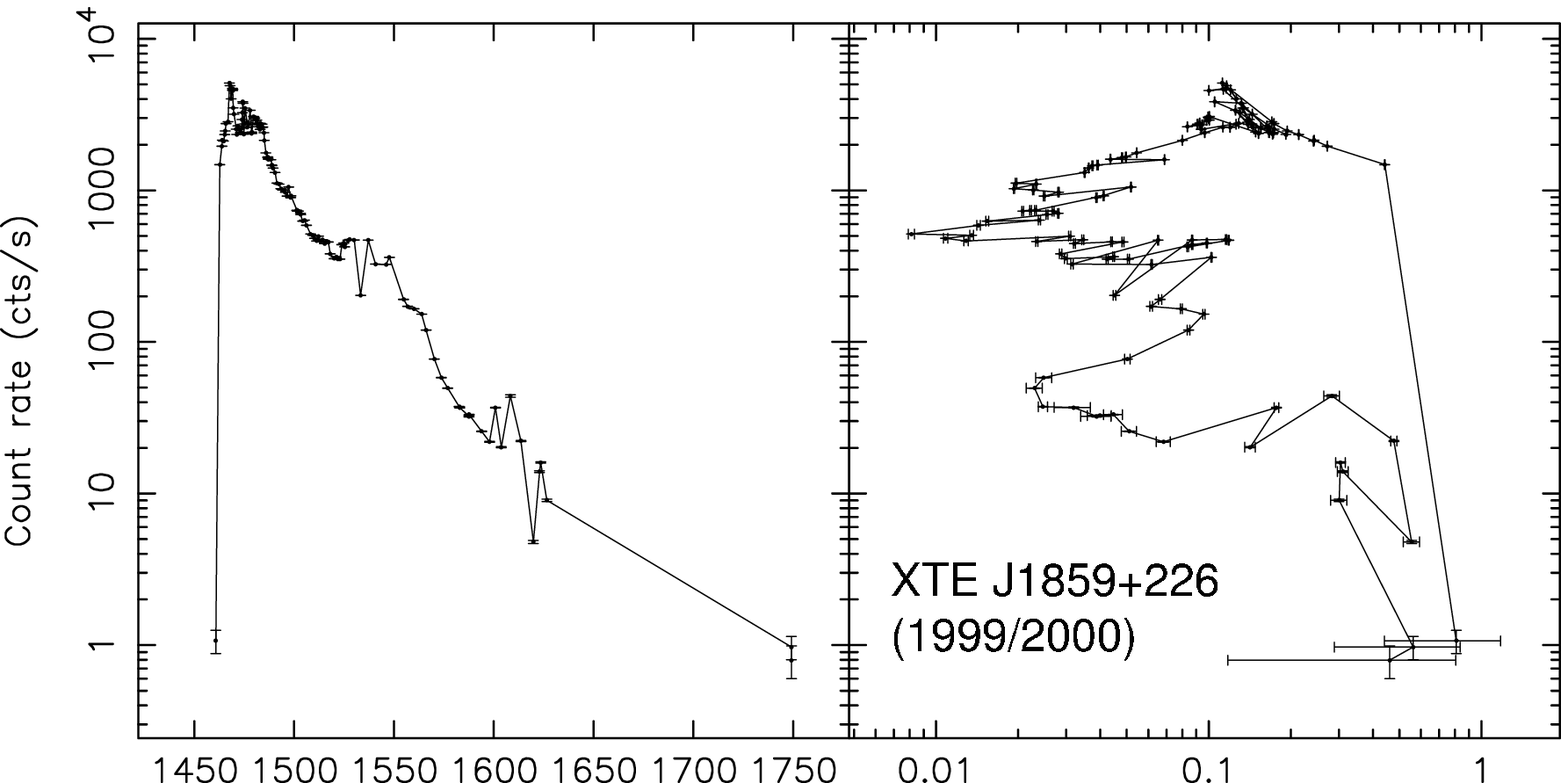}\\
		\includegraphics[height=6.54cm]{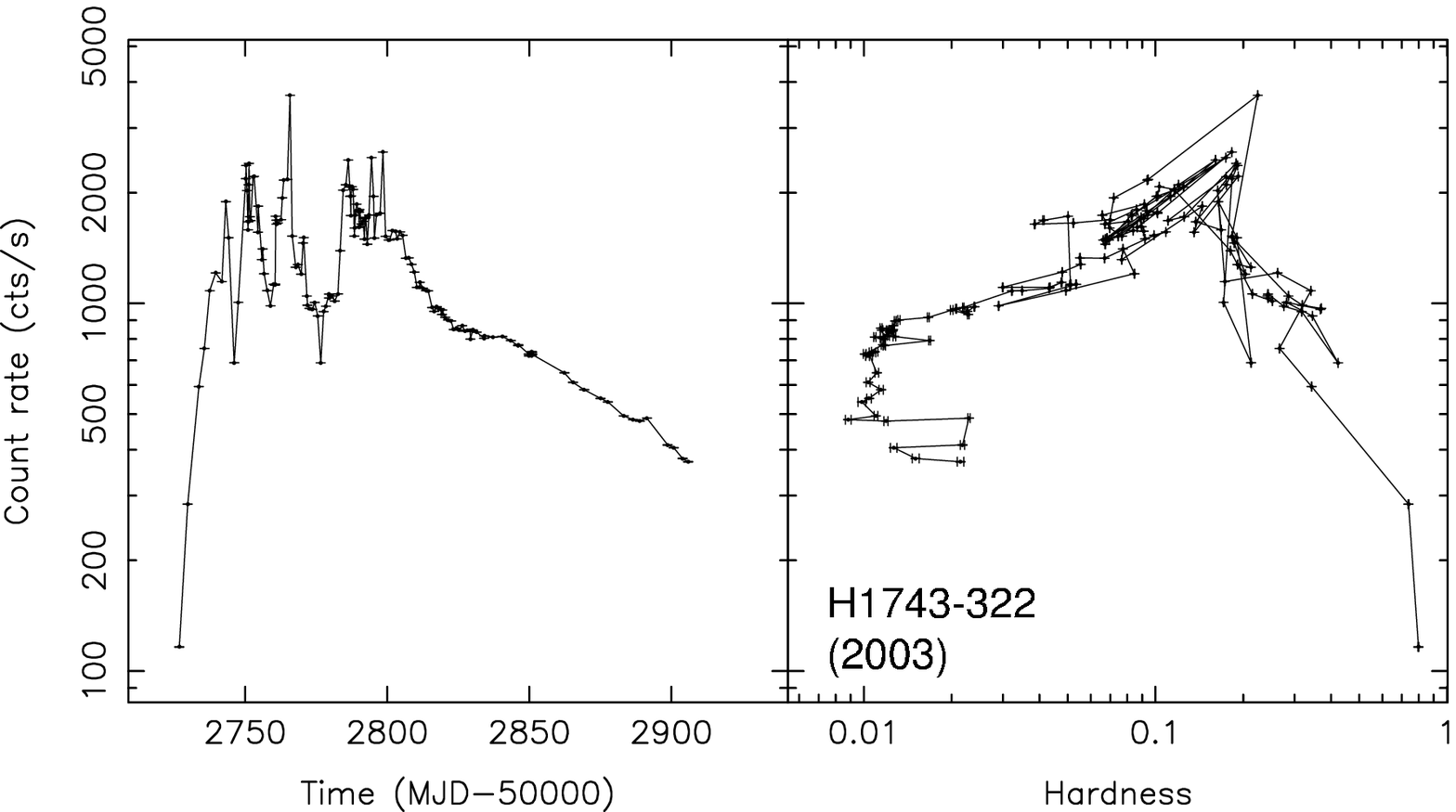}
   \end{tabular}
\caption{
Light curves (left) and HIDs (right) of three transients observed with the RXTE/PCA (from \cite{HomBel05}).
}
\label{fig:hombel05}
\end{figure}

The diagrams of all other BHT observed by RXTE, although the shape of the HID deviate from the `q' shown here, are qualitatively in agreement with this picture \cite{HomBel05,HomBel08,FenderHomanBelloni09}. 
Three examples can be seen in Fig. \ref{fig:hombel05}.

The general time sequence of the points in the HID is clear: the `q' diagram is followed in a counter-clockwise direction, although there are clearly more complex movements on the left branch. This will be discussed below in Sect. \ref{sec:4}.

\section{Aperiodic variability}
\label{sec:3}

The next step is to examine the properties of fast aperiodic variability (see \cite{vdkbook}). I will concentrate on the Power Density Spectra (PDS): although important information can be extracted from higher-order timing tools, they are not essential for the basic determination of states and state transitions.

\subsection{Classification of Power Density Spectra\index{Power density spectrum}}

\begin{figure}
\centering
\includegraphics[height=9cm]{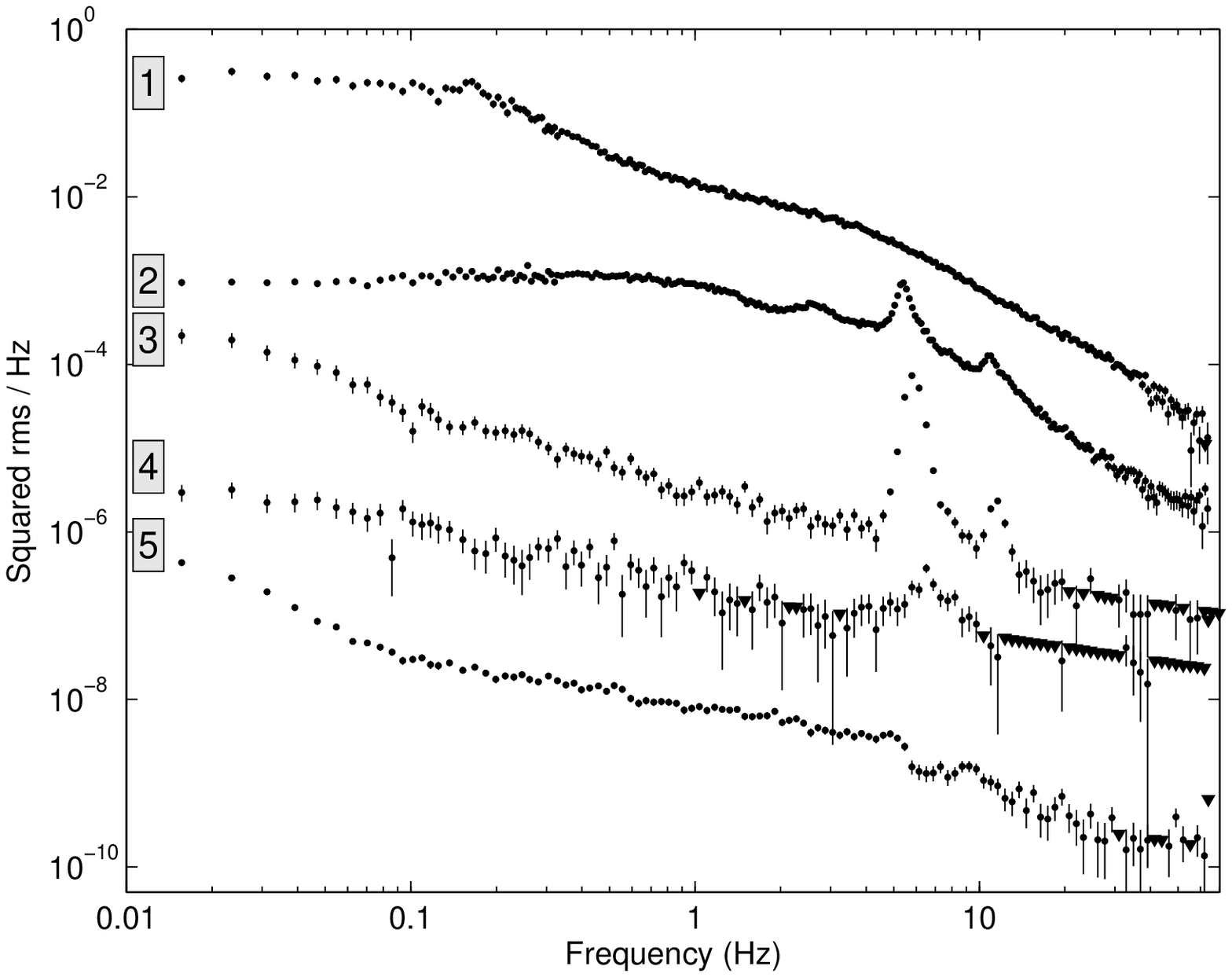}
\caption{
Examples of the five types of PDS described in the text (labeled on the left). The data are from the 2002/2003 outburst GX~339-4 \cite{Belloni05} and correspond to the energy range 3.8--21.2 keV. The PDS are shifted in power for clarity
}
\label{fig:pds339}
\end{figure}

The original paper on GX~339-4 reports a number of different PDS shapes, but it is now clear that we can classify them as belonging to few basic types (see \cite{HomBel08,CasellaQPO,Casella1859}). Furthermore, as we will see below, these types are closely related to the position on the HID and HRD. All the basic types but two were observed in the PDS of the 2002-2003 outburst of GX~339-4: they are shown in Fig. \ref{fig:pds339}.

\begin{itemize}
\item
PDS labeled 1 in Fig. \ref{fig:pds339} corresponds to the hard points in the HID. Its shape can be fitted with a small number, three to four, of very broad Lorentzian\index{Lorentzian} components, plus in some case a low-frequency \index{QPO}QPO peak \cite{Nowak2000,bpk,Belloni05}. The characteristic frequencies of all these components increase with source flux, while since the HID branch is almost vertical, the hardness remains almost unchanged, with only a minor softening at high flux. As it can be seen in the HRD, the total fractional rms on the right branch is rather high, larger than 20\%. As a function of energy, the fractional rms shows a slight decrease (see e.g. \cite{Gzdz05}). This is the band-limited noise typical of the LHS \cite{Nowak2000,Pottschmidt2003}.

\item
PDS 2 in Fig. \ref{fig:pds339} can be considered a high-frequency extension of PDS 1 and can be found at intermediate hardness values, notably along the horizontal branches. It can be decomposed in a number of Lorentzian components which correspond to those found in the LHS (see e.g. \cite{Belloni05}). The most prominent feature is a QPO with centroid frequency varying between $\sim$0.01 and $\sim$20 Hz. All Lorentzian components vary together, including the QPO. Unlike the previous case, they are strongly correlated with hardness: softer spectra correspond to higher frequencies and also to lower integrated rms variability (see Fig. \ref{fig:hid339}).
The low-frequency QPO (LFQPO) observed here is termed ``type-C'' QPO\index{type-C QPO} (see \cite{CasellaQPO} for precise definitions of QPO types): its most important property for our description is that it always appears together with moderately strong ($\sim$5--20\% rms) band-limited noise. The total fractional rms \emph{increases} with energy, in marked difference with PDS 1 \cite{Gzdz05}.
It is usually accompanied by at least two peaks harmonically related: one at half the frequency and one at twice the frequency. Although the QPO frequency is correlated with spectral hardness, on short ($<$10 s) time scales, it is rather stable.

The frequency of the type-C QPO is very strongly correlated with the characteristic (break) frequency of the underlying broad-band noise components \cite{wvdk,bpk}, a correlation which also extends to neutron-star binaries (see \cite{vdkbook}). The two main band-limited noise components have break frequencies which are one at the same frequency of the QPO and the other a factor of five lower \cite{wvdk,bpk}. Often, only one  broad-band component is detected, depending on the energy band considered, as their energy dependence is rather different (see Fig.~4 in \cite{Cui1550}).

\item
PDS 3 in Fig. \ref{fig:pds339} (found over a rather narrow range of intermediate hardnesses) also shows a QPO (called ``type-B'' QPO\index{type-B QPO}), but with very different characteristics than the one discussed above.
A detailed discussion of the difference between QPO types can be found in \cite{CasellaQPO}. The fact that type-B  QPOs are not simply an evolution of type-C QPOs is demonstrated by the fast transitions observed between them (see \cite{Casella1859} and by the few cases of simultaneous detection (i.e. GRO~J1655-40 observed by RXTE on 2006 May 26).
The total fractional variability is lower, due mainly to the fact that the band-limited components are replaced by steeper and weaker components. As for PDS 2, the fractional rms increases with energy. Type-B QPOs show a harmonic structure similar to that of type-C QPOs (two harmonic peaks are visible in Fig. \ref{fig:pds339}, while often a peak at 1/2 the frequency of the main peak is observed). While type-C QPOs span  large range in frequency, type-B QPOs are limited to the range 1--6 Hz, but detections during high-flux intervals are concentrated in the narrow 4--6 Hz range \cite{CasellaQPO}. The centroid frequency appears positively correlated with source intensity rather than hardness (as they are associated to a very narrow range of hardness, see below).

A closer look to Fig. \ref{fig:pds339} will reveal that the shape of the type-B peak is different from that of type-C. While the latter can be fitted with a Lorentzian model, type-B QPOs are consistent with a Gaussian shape, with broader wings. This is due to the fact that type-B QPOs jitter in time on short time scales, to the effect that in the average PDS its peak is smoothed \cite{Nespoli03}. This can be seen in Fig. \ref{fig:nespoli}, where a spectrogram of six minutes of PCA observation of GX~339-4 is shown: the jitter in frequency takes place on a time-scale of 10 seconds \cite{Nespoli03}.

\item
The last PDS in Fig. \ref{fig:pds339} showing a QPO is \#4, found at hardness values systematically slightly lower than those of PDS 3. Here, a so-called ``type-A'' QPO\index{type-A QPO} is shown. Being a much weaker and broader feature, we know less details about this oscillation; sometimes it is only detected by averaging observations. Its frequency is always in the very narrow range 6--8 Hz and it is associated to an even lower level of noise than type-B. In fact, the three types of QPO can be separated by plotting them against the integrated fractional rms of the PDS in which they appear \cite{CasellaQPO}.

\begin{figure} 
\centering
\includegraphics[height=9cm]{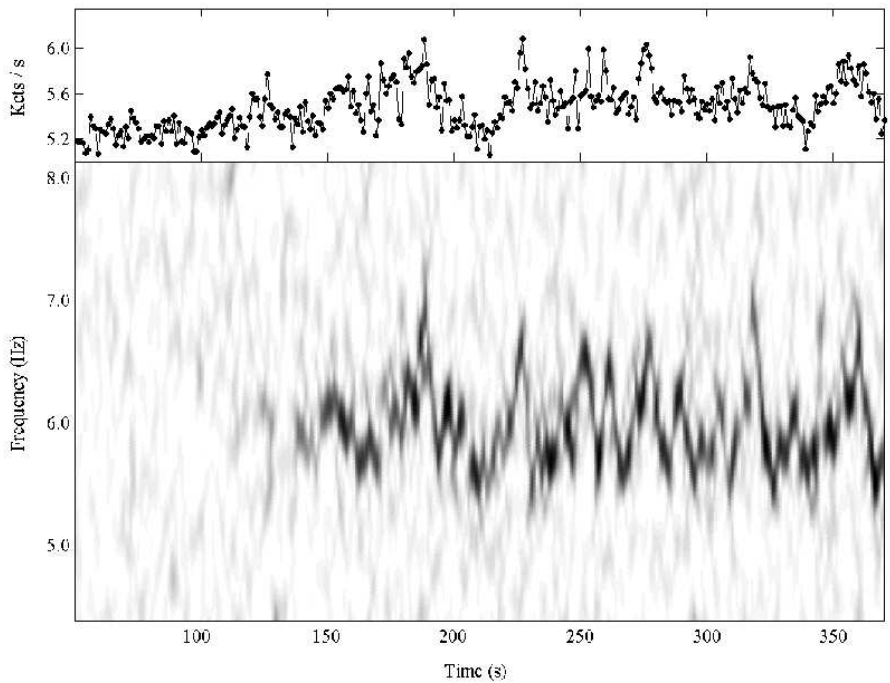}
\caption{The first six minutes of the PCA observation of GX~339-4 from 2002 May 17. Top panel: light curve with 1-second binning. Bottom panel: corresponding spectrogram in the 4--8 Hz range; darker regions correspond to higher power. The onset of a sharp and strong type-B QPO at around 6 Hz is evident, as well as the variability of its centroid frequency. In the first two minutes, a type-A QPO is present, but it too weak to show in the spectrogram. From \cite{Nespoli03}
}
\label{fig:nespoli}
\end{figure}

\item
Finally, PDS 5 in Fig. \ref{fig:pds339} shows a weak steep component, which often needs a long integration time for a detection. Weak QPOs at frequencies $>$10 Hz are sometimes observed, as well as a steepening/break at high frequencies. The total fractional rms can be as low as 1\%, increasing with energy \cite{Gzdz05}. This PDS corresponds to the soft points at the extreme left of the HID.

\begin{figure}
\centering
\includegraphics[height=9.5cm]{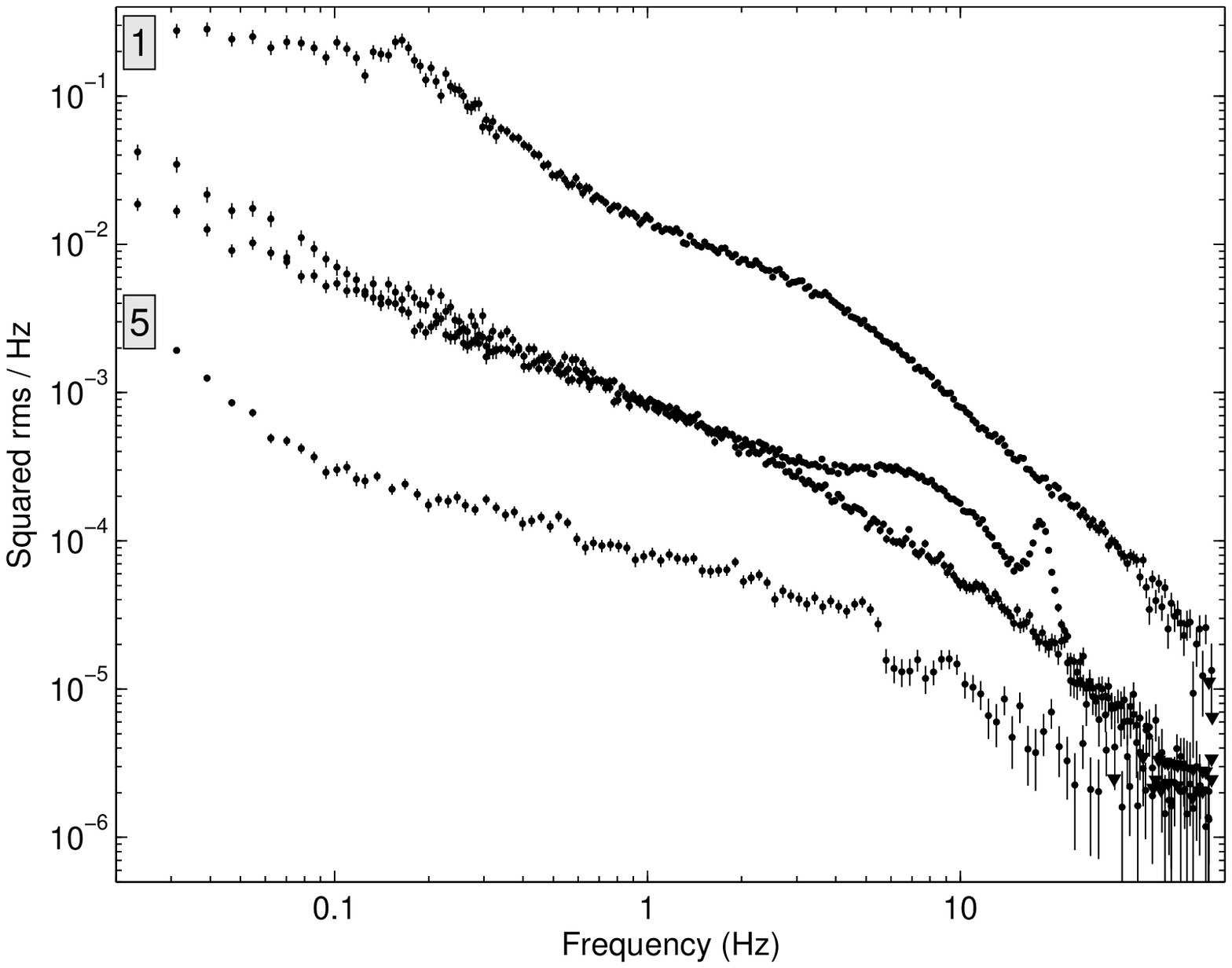}
\caption{Examples of the two types of PDS which have not been found in GX~339-4 (see text). The top and bottom PDS are \# 1 and 5 from Fig. \ref{fig:pds339}, shifted in power for clarity. The middle ones correspond to two observations of GRO~J1655-40\index{GRO~J1655-40} from its 2005 outburst: the one with a QPO from 2005 May 24, the other from 2005 May 8. The middle ones are not shifted in power
}
\label{fig:pds1655}
\end{figure}

\item
As mentioned above, in addition to the PDS shapes in Fig. \ref{fig:pds339}, there are two types of PDS which was not observed in GX 339-4. Examples from the 2005 outburst of GRO~J1655-40 can be seen in Fig. \ref{fig:pds1655}, together with PDS 1 and 5 from Fig.  \ref{fig:pds339}. The integrated fractional rms of both is intermediate. One PDS has a featureless curved shape, the other is similar below a few Hz, but shows an additional bump and a QPO. As we will see, these ``anomalous'' PDS shapes are associated to anomalous ``flaring states''.

\end{itemize}

\subsection{Fast transitions}

The PDS described in the previous subsection have been obtained as averages over hundreds or thousands of seconds of observations. While for most cases this averaging procedure is justified, there are observations where very fast transitions are observed, requiring a time-resolved analysis. An example is shown in Fig. \ref{fig:nespoli}. Not only the centroid frequency of the QPO jitters with time, but the oscillation is clearly not visible in the first two minutes of data. Averaging the first part, a type-A QPO appears \cite{Nespoli03}. These transitions are common and have been observed with RXTE in many sources \cite{Belloni05,Casella1859,HomBel08}: they correspond to those already discovered with Ginga \cite{Miyamoto94,Takizawa97}. All of them, with no exception, involve a transition between a type-B and a type-A QPO, or between a type-B and a type-C QPO. A number of detailed transitions from XTE~J1859+226 can be found in \cite{Casella1859}: there, an exponentially time-decaying threshold in count rate was found for the occurrence of the three types of QPO. This is clearly not a common case, since from Fig. \ref{fig:nespoli} it is evident that the count rate corresponding to the type-A QPO for GX~339-4 is lower than that of the type-B QPO, opposite to the case of XTE~J1859+226.

\section{The time evolution}
\label{sec:4}

\begin{figure}
\centering
\includegraphics[height=9.5cm]{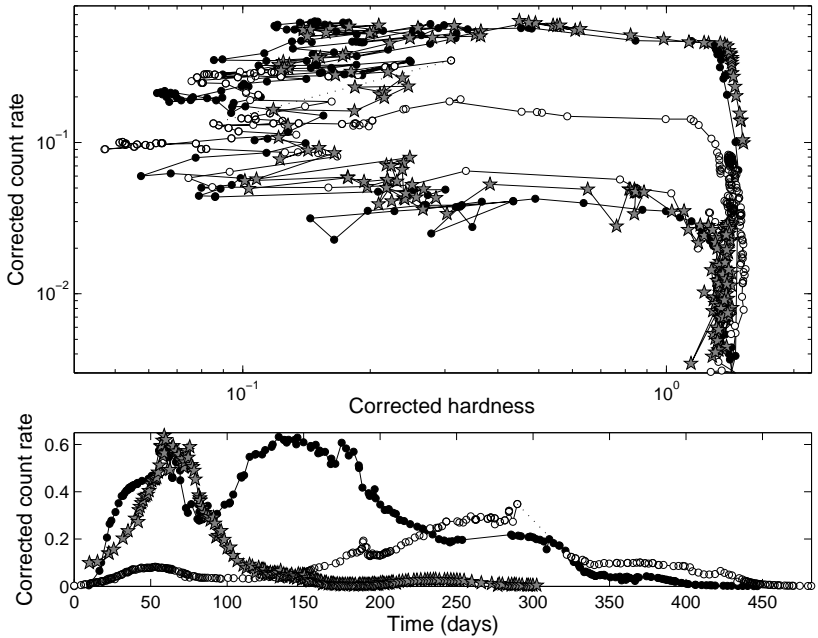}
\caption{
Top panel: HIDs for the three recent outbursts of GX~339-4. In order to compare different epochs the spectra have been divided by a simulated Crab spectrum for each observation date. Therefore, the count rate is in Crab units in the 3.8--21.2 keV band and the hardness of 1 corresponds to a Crab spectrum. The symbols identify the three outbursts (2002/2003: filled circles; 2004/2005: empty circles; 2007: gray stars). The dashed line indicate a long time gaps. 
Bottom panel: corresponding light curves plotted on the same scale. Symbols are the same
}
\label{fig:comparison_339}
\end{figure}

In what presented above, the time evolution of the X-ray properties was not shown, as the attention was focused on the different classes of hardness and variability properties. The light curves of transient BHB can be quite diverse (see e.g. \cite{HomBel05,McRembook,HomBel08}). GX~339-4 has had three major outbursts in the past five years. In the bottom panel of Fig. \ref{fig:comparison_339} one can see that the time evolution of the three events is very different. 
Despite these differences, a general behavior can be identified, leading to basic facts shared by all sources. The top panel of Fig. \ref{fig:comparison_339} shows that the three outbursts of GX~339-4 had a very similar evolution, despite the differences in the time domain. Other sources show a HID qualitatively similar to the q-shaped diagram of GX~339-4 (see \cite{HomBel05} and Fig. \ref{fig:hombel05}), while others behaved more erratically (see \cite{McRembook,HomBel08}).

The basic properties of the time evolution of outbursts are the following:

\begin{itemize}

\item All sources start from quiescence, become bright for a period typically of several months, then faint again and return to quiescence (with the exception of the three persistent bright BHB Cyg~X-1, LMC~X-3 and LMC~X-1, and of GRS~1915+105 which is still very bright after fifteen years). This is the definition of transient X-ray sources.

\item For some sources, the initial rising part of the outburst was not observed (see for instance XTE~J1650-500 in Fig. \ref{fig:hombel05}). The first pointed RXTE observations already show a high flux level. All sources are observed in the later parts of the outburst to become faint down to the minimum detectable levels with the RXTE PCA. This is obviously due to the fact that an unknown source or outburst must be discovered, leading to a delay in pointed observations. However, the typical reaction time is at maximum a few days. This means that the initial rise can be as fast as a couple of days in these sources.

\item In all sources for which the initial rise is observed, the start of the outburst is hard and so is the end of the outburst (see Fig. \ref{fig:comparison_339}). There is a number of sources for which pointed observations started late and that were initially observed with a low hardness. Since the initial rise was fast, this is compatible with a hard rise.

\item Most sources do show significant changes in hardness, extending to the left branch in the HID, corresponding to a soft spectrum. Intermediate values of hardness are observed. A few sources, such as XTE~J1118+480, did not leave the hard branch and move in the HID on a roughly vertical hard track (see \cite{Filippo1118,Mc1118}). This also applies to some outbursts of BHB which have shown other outbursts with a complex structure. An example is XTE~J1550-564: its first two outbursts were complex and showed spectral variations (see \cite{Homan1550,Jerome1550}), while the next (fainter) two did not (see \cite{Belloni1550,Sturner1550}).

\item A good number of sources followed an evolution similar to that of GX~339-4 (Figs. \ref{fig:hombel05} and \ref{fig:comparison_339}): the q-shaped pattern was followed in counter-clockwise direction. While there can be many state transitions, the q shape identifies two main ones: a hard-to-soft one at high flux in the first part of the outburst, and a soft-to-hard one at the end of the outburst (see below). The first one is particularly important since it has been shown to be associated to main ejections of relativistic jets (see \cite{fbg04}).

\item From the analysis of data from the Ginga satellite, the presence of a hysteretic behavior in the evolution of transient BHB was noticed \cite{Miya_hysteresis}: the transition from the right to the left branch, namely from hard to soft, takes place at higher luminosities than the reverse transition later in the outburst. This was also found with RXTE data of transient BHB and the neutron-star transient Aql~X-1 \cite{Maccarone_coppi} and is evident from Fig. \ref{fig:comparison_339} for the case of GX 339--4. Other cases can be seen in \cite{Homan1550,HomBel05, RemMcARAA, HomBel08}. In the case of GX~339-4, which showed three outbursts qualitatively similar to each other, it appears as the luminosity difference between the bright and the faint transition (the two horizontal branches in the HID) is related to the level of the first.
It was suggested that the flux at the initial hard peak (roughly corresponding to the top of the right branch in the HID) is correlated with the time from the final hard peak (the return to the LHS at the end of the outburst) of the previous outburst \cite{yu_peaks}. This correlation is also followed by the 2007 outburst.
\end{itemize}

\section{Definitions of source states\index{states}}
\label{sec:5}

From what discussed in the previous sections, we can identify a number of source states that must be examined separately. As shown above, many observed properties change smoothly throughout the basic diagrams, but some do not. In particular, it is the inspection of the fast-variability properties which indicates the presence of abrupt variations. It is these sharp changes that must be taken as landmarks to separate different states. Whether these correspond to actual physical changes in the accretion flow is obviously not a priori clear and must be determined with the application of models. This approach has the advantage of being phenomenological and completely model-independent and is meant to provide theoreticians with a solid observational framework on which to base their investigation. 
Below, I present a revised state classification, together with the observed transitions between them. This classification is similar but not identical to that presented in \cite{HomBel05}. A graphical summary can be seen in Fig. \ref{fig:states}.

\begin{itemize}

\item{\it Low-Hard State (LHS):} it is associated \emph{only} to the early and late phases of an outburst and it corresponds to the vertical branch at the extreme right of the HID. At the end of an outburst, once this state is reached, no more transitions have ever been observed. In the spectral domain, it is characterized by a hard spectrum, with a power-law index of 1.6--1.7 (in the 2-20 keV band) and little moderate variations (right-most zone in the HID). A high level of aperiodic variability is seen in the form of strong band-limited noise components (PDS 1 in Fig. \ref{fig:pds339}), with typical rms values of $\sim$30\%, anti-correlated with flux and positively correlated with hardness. The PDS can be decomposed in a number of Lorentzian components (see \cite{Nowak2000,bpk,Pottschmidt2003,MarcMichiel}), whose characteristic frequencies increase with flux. One of these components can take the form of a type-C QPO peak. The time spent in this state, both at the beginning and at the end of the outburst, can be quite variable. 

\begin{figure}
\centering
\includegraphics[height=13cm]{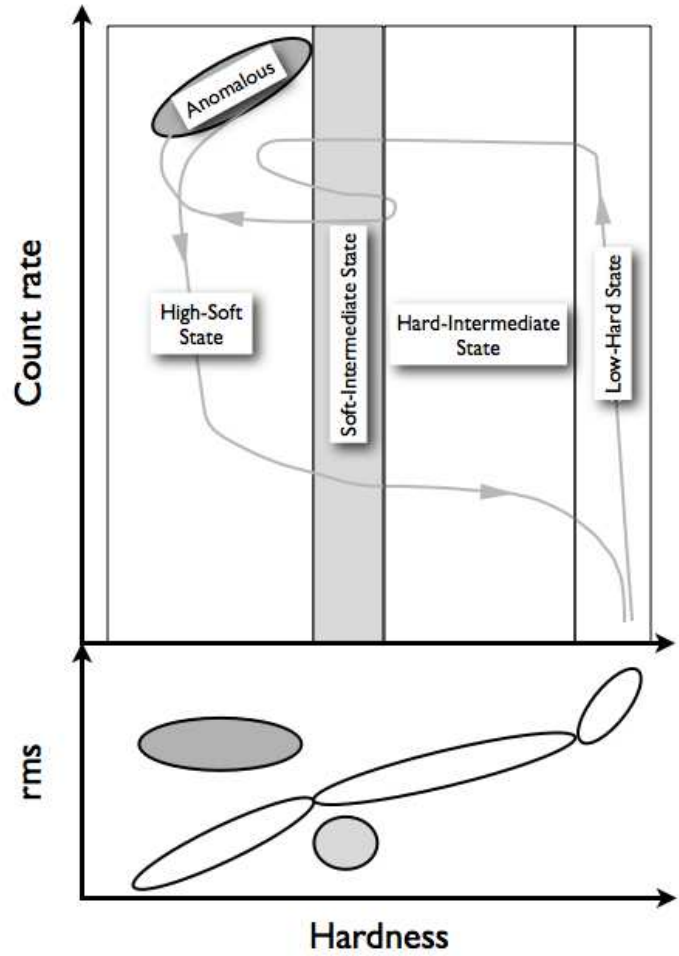}
\caption{
Sketch of the general behavior of a black-hole transients in the HID (top) and HRD (bottom), with the HID regions corresponding to the states described in the text.
}
\label{fig:states}
\end{figure}

\item{\it Transition to the Hard-Intermediate State:} from the HID and HRD, it is difficult to mark a precise end of the LHS and the start of the HIMS. As shown in the next bullet, the properties of the HIMS are compatible with being an extension of those of the LHS. Changes in the speed of increase of the timing components, or in the frequency-frequency correlations can be seen (\cite{Belloni05}), but the clearest marker for the transition has been observed in the IR/X-ray correlation for GX~339-4: the change in the  correlation was very fast and marked and provided a precise marker for the transition \cite{HomanIR}. Without making use of observation at other wavelength, the  time of the transition can only be identified with an uncertainty of a few days.

\item{\it Hard-Intermediate State (HIMS):} at the start of the outburst, if the source ever leaves the LHS, it enters the HIMS, moving along the top horizontal branch in the HID. The spectrum is softer, the softening being due to two simultaneous effects: an increase of the power-law index that can reach 2.4--2.5 and the appearance of a thermal disk component. The fast aperiodic variability corresponds to PDS B in Fig. \ref{fig:pds339}, with a band-limited noise and a strong type-C QPO. The total fractional rms is lower than in the LHS (10--20\%) and decreases with softening. The  PDS can be decomposed in the same Lorentzian components as in the LHS, with characteristic frequencies that are higher than in the LHS. In the LHS, the frequencies increased with flux, with little spectral variations, in the HIMS they increase with softening of the energy spectrum. The top branch in the HID is traveled rather fast and lasts at most a dozen days.
This state is also observed in the central and final parts of the outburst (see below). 

\item{\it Transition to the Soft-Intermediate State (Jet line):} this transition is marked \emph{only} by the timing properties. The overall level of noise drops (see HRD) and a type-B QPO appears in the PDS (PDS 3). All this takes place with a minor softening of the spectrum: this abrupt transition associated to a small change in hardness is associated to the ``jet line'' in the HID \cite{Belloni05,fbg04}. The time coincidence of the crossing of he jet line and the HIMS-SIMS transition is however not exact \cite{FenderHomanBelloni09}.

\item{\it Soft-Intermediate State (SIMS):} although spectrally below 20 keV the energy spectrum is only slightly softer than in the HIMS, as shown by the similar hardness, the timing properties are radically different and mark a clearly different state. The lower level of variability, which can be as low as a few \% is a clear indication of this state (see the HRD). A QPO is often present, either of type A or of type B (PDS 3 and 4 in Fig. \ref{fig:pds339}). It is not clear whether it is always present, since observations without clear QPO peaks could feature an undetected type-A QPO. 

\item{\it Transitions from and to the Soft-Intermediate State:} once the SIMS is reached, a number of transitions can be seen. All of them involve moving from or to the SIMS. In Fig. \ref{fig:states} this is exemplified with a short return to the HIMS, but the situation can be much more complex. Very fast transitions such as that shown in Fig. \ref{fig:nespoli} can be observed, all of them involving a type-B QPO (see also \cite{Casella1859}). In particular, the jet line can be crossed more than once.

\item{\it High-Soft State (HSS):} this soft state corresponds to the softest spectrum, dominated by a thermal disk component, with only a small contribution to the flux by a hard component. The variability is in the form of a weak (down to 1\% fractional rms) steep component. Weak QPOs are sometimes detected in the 10--30 Hz range. The hardness variations shown in Fig. \ref{fig:hid339} are due to changes in the hard component.
 
\item{\it ``Anomalous'' State:} In some sources, a different class of PDS has been seen at high flux (see Fig. \ref{fig:pds1655}). Although the hardness corresponds to that of the HSS (and sometimes to that of the SIMS), there is evidence that the energy spectrum is different (see below). The anomalous PDS are also accompanied by a higher integrated variability (see HRD).

\item{\it Transitions back to the Low-Hard State:} The HIMS presents itself again at the end of the outburst, with very similar properties, preceding the final LHS. The timing properties along this transition are smoother, also because of the lower statistics associated to the lower flux of these observations. 
The range in hardness of the HIMS is similar at low fluxes, but not identical, as the SIMS appears to cover a smaller extent in hardness.
These transitions during outburst decay are extensively presented in \cite{Kalemci2004} and references therein.

\end{itemize}

As mentioned above, the source states outlined above stem from the hardness/timing properties observed in all systems observed with RossiXTE. They do not necessarily correspond to markedly different physical conditions. Indeed, the only transition which marks sharp changes in the timing properties is the HIMS--SIMS one. Moreover, looking at the HRD, it appears that LHS, HIMS and HSS follow a single path, suggesting that they have something in common. As mentioned above, the HIMS can be seen as an extension of the LHS, but also the HSS shows similar timing properties (a low-frequency QPO, aperiodic noise decreasing as the source softens). The only states which are radically different are the SIMS and the anomalous one. In Sect. \ref{sec:8}, we will see what the situation is in terms of broad-band emission and spectral models.

\subsection{Comparison with other state classifications\index{states}}

Although different authors use different terms, even mixing nomenclatures, there are only two other state classification in addition to the one presented here. It is important  to compare them in order to clarify similarities and differences.

\subsubsection{Ginga canonical states}

The original `canonical' states introduced on the basis of observations made by Ginga were the following. A hard (called low) state and a soft (called high) state were defined in a similar fashion as was done before. The hard state \cite{MiyamotoCanonical} was identified with strong variability in form of a band-limited noise whose break frequency is correlated with the flat-top level (see \cite{BelloniHasinger,Mendez97}), with hard phase lags increasing with energy and a 2-20 keV energy spectrum characterized by a power law with photon index $\sim$1.5. The soft state by a strong thermal accretion disk energy spectrum associated to a very low level of aperiodic variability. The new ``very-high'' state (VHS) was originally identified in the brightest observations of GS~1124-684 and GX~339-4 on the basis of a different PDS, with lower variability, higher characteristic frequencies and a QPO \cite{MiyamotoCanonical}. Subsequently, a second ``flavor'' of VHS was reported, with a different PDS shape and a lower integrated fractional rms \cite{MiyamotoAnother}. This was obviously associated to a different type of QPO \cite{Takizawa97}. Finally, all the characteristics of the original VHS were discovered in both GX~339-4 and GS~1124-684 at much lower fluxes, observed after a long interval of soft state. This led to the tentative inclusion of an intermediate state taking place between the soft and the hard states \cite{Belloni97,Mendez97}. 

Given the relatively sparse time coverage obtained by Ginga, it was not possible to follow in detail transitions between states (with the exception of fast ``flip--flops'', which were not recognized as transitions \cite{Miyamoto91}). This prevents a precise comparison. However, in addition to the obvious identification of the LHS and HSS, the two VHS flavors can be identified with the HIMS and the SIMS. The need for a separate intermediate state was removed by the discovery that mass accretion rate is not the only parameter driving the evolution of the outbursts \cite{Homan1550}.

\subsubsection{Quantitative state classification}

More recently, a state classification based on the determination of timing and spectral parameters has been proposed (see \cite{McRembook,RemMcARAA}). The underlying idea is to have a definition based on instrument-independent parameters, with a precise definition of states which makes possible a comparison between sources. An updated definition of states can be found in \cite{Mc1743}. Three states are presented: Hard, Thermal and Steep Power Law (SPL), which are identified with the LHS, HSS and VHS of the Ginga classification respectively. Their definition is based on precise boundaries of a number of parameters such as integrated fractional rms and presence of a QPO in the PDS, power-law photon index and disk fraction in the energy spectra. These three states do not fill the complete parameter space and all observations which do not qualify are generically classified as ``intermediate''. 
Among the RXTE observations of the 2003 outburst of H~1743-322, about 17\% of the observations were intermediate between two states. A large fraction of observations of 4U~1630-47 also did not fit the three-state classification \cite{Tomsick1630}.
Therefore, this classification is not meant to be exhaustive like the one presented above, but rather to guide through general parameters of emission, such as flux ratio between main spectral components, spectral indices and fractional rms of QPOs. Whether to adopt one or the other is dependent on what one's final aim is.

As mentioned above, the LHS and HSS are normally undisputed (but see below for the HSS): what is difficult to identify is their boundary. As I have shown in Sect. \ref{sec:5}, there are sharp changes in some observable which can be taken as markers of state transitions (IR/X correlation for LHS-HIMS, rms drop for other transitions). It is not clear at this stage whether the boundaries in this classification can reproduce these transitions.   In particular, the SPL is defined in a rather complex way which involves the presence of a QPO of either type and the presence of a power-law component with photon index $>$2.4. The HIMS-SIMS transitions of GX 339-4 were observed at different power-law indices: in 2002 the index was 2.44 before \emph{and} after the transition \cite{Nespoli03}, in 2004 is varied from 1.9 to 2.3 \cite{Belloni_INTEGRAL}. It appears as this transition falls well within a single state. On the other hand, the spectral analysis of five years of RXTE observations of Cyg~X-1 showed that a photon index of 2.1 indeed marks a transition between the LHS and a softer state \cite{Wilms06}.

As to the comparison with the system presented here, independent of the precise parameters adopted for the states, which have evolved with time, it is clear that the Hard and Soft can be identified with the LHS and HSS respectively. The SPL and the different intermediate states would then correspond (roughly) to the SIMS and HIMS. 

\section{High-frequency QPOs\index{QPO}}
\label{sec:6}

Particular importance is attached to the high-frequency ($>$30 Hz) QPOs (HFQPO) detected in the PDS of some BHB. There are few instances of such detections, yet the situation here is entangled and even though few publications are available, it is difficult to derive a clear pattern. Here I summarize the basic information available, relying on significant detections of narrow (Q$>$2) features with a centroid frequency above 20 Hz (references can be found in \cite{Belloni_hfqpo}). Single QPO peaks have been detected from three sources: XTE~J1650-500 (250 Hz), 4U 1630--47 (variable frequency) and XTE~J1859+226 (90 Hz).  
From four other sources, pairs of QPO peaks have been detected: GRO~J1655-40 (300/400 Hz),
XTE~J1550-564 (184/276 Hz),
H~1743-322 (165/241 Hz)  
and GRS~1915+105 (41/69 Hz). 
The case of GRS~1915+105 is more complex and will be discussed below. All detected peaks are weak, with a fractional rms of a few \% and strongly dependent on energy. In some cases, only a few detections are available, but in others many peaks have been found: the centroid frequencies are not always constant, as one can see in the case of XTE~J1550-564 and 4U~1630-47. However, it has been first noticed by \cite{AbrKlu} that when two simultaneous peaks are detected, their frequencies are in a 3:2 ratio in three cases out of four, while for GRS~1915+105 the ratio is 5:3. It is interesting to note that for XTE~J1550-564, a double peak was discovered averaging a few observations from the 2000 outburst \cite{Miller2001qpo}, while for the previous outburst in 1998 many single detections were reported \cite{Remillard1550}: those associated to type-B QPOs cluster around 180 Hz, while those associated to type-A QPOs are around 280 Hz.
The fact that some sources appear to have preferred frequencies suggests that they are associated to basic parameters of the black hole. Indeed, a degree of anti-correlation with the dynamically-dermined black-hole mass has been found (see \cite{RemMcARAA} but also \cite{Belloni_hfqpo}).

The comparison with kHz QPOs in neutron-star X-ray binaries shows that these are most likely different features: unlike the BHB HFQPO, which are mostly observed at the same frequency and have never been detected together with type-C WPOs, kHz QPOs span a large range in frequency and are correlated with the low-frequency QPOs which are the NS counterpart of type-C (see \cite{vdkbook,bpk}). Moreover, the kHz QPOs have been shown not to have preferred frequencies \cite{bmh1,bmh2}

For this paper, we are interested in where observations with an HFQPO are located in the HID and HRD and to which PDS shape they are associated. From the literature, many HFQPOs correspond to observations in the SIMS, i.e. associated to power-law noise and often type-A/B QPOs. There exist a a few exceptions, notably the case of XTE~J1550-564, where one detection was reported together with a type-C QPO \cite{Homan1550}; in this case, the frequency of the type-C QPO was high, indicating that the source was close to the SIMS.

\section{Other sources}
\label{sec:7}

The paradigm presented above has been derived from black-hole transients. These constitute the large majority of known black-hole binaries, but there are a few other systems need to be examined. There are persistent systems, such as Cyg~X-1, LMC~X-1 and LMC~X-3. Because of their large distance, the latter two are weak sources in our instruments, which means that the statistical level of the signal is much lower. LMC X--1 appears to be locked in the HSS, as does LMC X--3, which however has shown brief transitions to a harder state, even so extreme as not to be detected with RossiXTE. It is difficult to say more about them \cite{Wilms01,Nowak2001,Gotz06}. Moreover, it is impossible to ignore the prototypical microquasar GRS~1915+105 (see \cite{FenderBelloni}) which, despite its peculiarity, needs to be compared with other systems. Here I present these two sources, Cyg~X-1 and GRS~1915+105 in a similar framework, analyzing the differences and similarities with those outlined above.

\subsection{Cygnus~X-1\index{Cyg~X-1}}

Cygnus~X-1 is not a transient system: it is found most of the time in the LHS (the original definitions of the states come from this source), with occasional transitions to a softer state, usually interpreted as a HSS, and a series of ``failed state transitions'' (see \cite{Pottschmidt2003,Wilms06}). However, already the first state-transition observed with RossiXTE in 1996 showed properties which were not compatible with those of a full-fledged HSS \cite{Belloni96}. The total integrated fractional rms is rather high even as the source softens and does not go down to a few \% as in transients (see \cite{Gleissner2004}). Spectrally, the hard component does become rather soft \cite{Wilms06}, but no sharp QPO like those shown above has been observed (see e.g. \cite{Pottschmidt2003}). In the energy spectrum, a change of properties has been identified corresponding to a low-energy photon index of 2.1 (see above), which was then taken as a marker of state transitions \cite{Wilms06}.

\begin{figure}
\centering
\includegraphics[height=11cm]{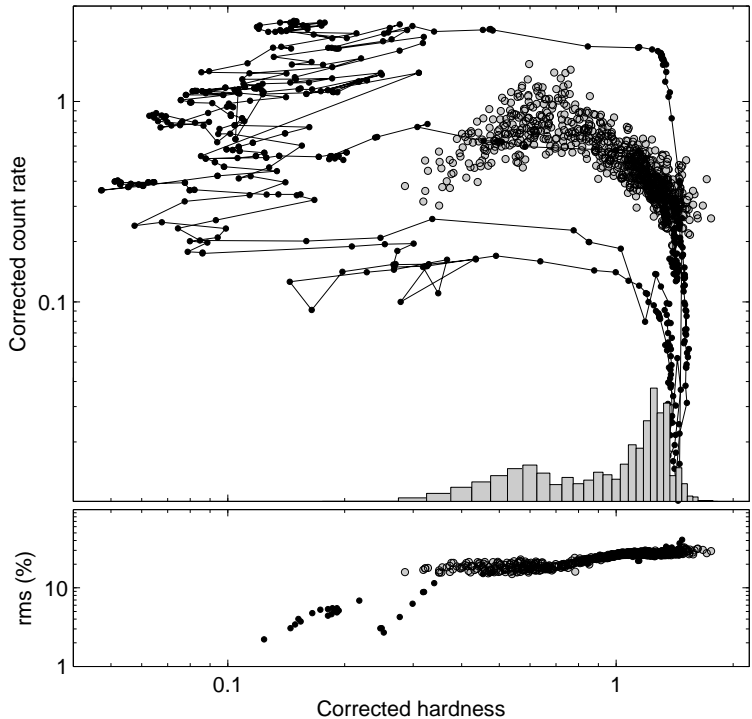}
\caption{
Top panel: HID of all the RXTE/PCA observation of Cygnus~X-1 (gray circles), plotted over the HID of the 2002/3 and 2004/5 outbursts of GX~339-4 (black circles). The gray histogram shows the distribution of hardness values, on the same X axis. The presence of two separate peaks is evident. Note that the GX~339-4 points have been shifted to bring the source to the same distance as Cygnus~X-1.
Bottom panel: HRD of the same observations, plotted over a few of the GX~339-4 from the 2002/3 outburst (see bottom panel of Fig. \ref{fig:hid339})
}
\label{fig:hid_cigno}
\end{figure}

We can compare the properties of Cyg~X-1 with those of transient systems by analyzing in the same way, through the production of HID/HRD. The result can be seen in Fig. \ref{fig:hid_cigno}, where the points corresponding to 1065 RossiXTE observations ranging from 1996 February  to 2005 October, plotted over the points from GX~339-4. The count rate here has been corrected also for the difference in distance between the two sources. Most of the observations found Cyg~X-1 in the LHS, as shown by the hardness histogram. Clearly, the histogram features a second peak at softer values, which can also be identified with a bend in the HID. However, Cyg~X-1 does not show type-C QPOs as would be expected on that branch. It is interesting to note that the SIMS region is reached only for a few observations, termed ``failed state-transitions'', but also no type-B QPOs were seen \cite{Pottschmidt2003}. In correspondence of most of these events, a small radio flare was observed \cite{Pottschmidt2003}.
The HRD shows a strong similarity with GX~339-4: the points are well correlated and match those for GX~339-4, besides the few soft points which do not show evidence for a rms drop. The rms difference between the first and subsequent LHS observations reported by \cite{Pottschmidt2003} is present but invisible in the HRD due to the number of points.
Interestingly, a transient relativistic jet has been observed from Cyg~X-1 \cite{FenderCyg}. It corresponds in time to one of the excursions to a soft spectrum, but only observed down to hardness $\sim$0.3 and not to the extreme values in Fig. \ref{fig:hid_cigno}.

Another major difference is the absence of observable hysteresis (see \cite{Maccarone_coppi}). Only a single branch is seen, which is traveled in both directions. However, GX~339-4 has shown that the weaker the outburst is, the smaller is the difference in flux between the hard-to-soft and soft-to-hard transitions (see Fig. \ref{fig:comparison_339}). As Cyg~X-1 never reaches very high accretion rate values, its behavior would be compatible with a very small unmeasurable level of hysteresis.
Clearly, this system has somewhat different properties, probably related to the fact that it is not a transient. Nevertheless, its behavior in the HID and the HRD is not very different from that of other sources.

\subsection{GRS 1915+105\index{GRS 1915+105}}

On the topic of microquasars, it is not possible not to discuss GRS~1915+105. This system shows very peculiar properties, which however can be successfully compared with those of other sources \cite{FenderBelloni}. Three separate states were defined for this source, with a naming convention that in this context appears particularly unfortunate: states A, B and C \cite{Belloni1915}. Both from the energy spectra and the PDS, it is clear that GRS~1915+105 never reaches a LHS: the high-energy end of the spectrum does not reach the typical 1.6 photon index, the thermal accretion disk is always present with a measured inner temperature $>$0.5 keV and a strong type-C QPO\index{type-C QPO} is present during state C. A comparative analysis of the PDS in the three states of GRS~1915+105 shows not only that the LHS is never reached, but also suggests that states A and B correspond to the ``anomalous state'' described above \cite{Pablo1915}. 

Until recently, only a type-C QPO was observed from GRS~1915+105 in its hardest state, state C. However, a time-resolved analysis of fast varying light curves has led to the discovery of a type-B QPO\index{type-B QPO} in correspondence to state transitions, confirming the association between this oscillation and spectral changes \cite{Paolo1915}.
It would be interesting to compare the HID and HRD of GRS~1915+105 with those of other more conventional transient sources in order to ascertain whether its peculiar states can be associated with the states outlined above and whether their evolution is compatible. The presence of type-C QPOs in state C suggests that it would correspond to the HIMS, while the softer states could be associated to the HSS. The fast ($<$1 second) transitions would include the SIMS, as shown by the detection of a type-B QPO. 
However, the typical choice of X-ray energy bands for the production of hardness fails for this system: in its A and B states, the thermal accretion disk component is so hot that it extends well into the high-energy band. As a result, the hardness of these states is higher than that of state C. Moreover, the fast transitions would require a large amount of additional analysis effort. 

\begin{figure}
\centering
\includegraphics[height=11cm]{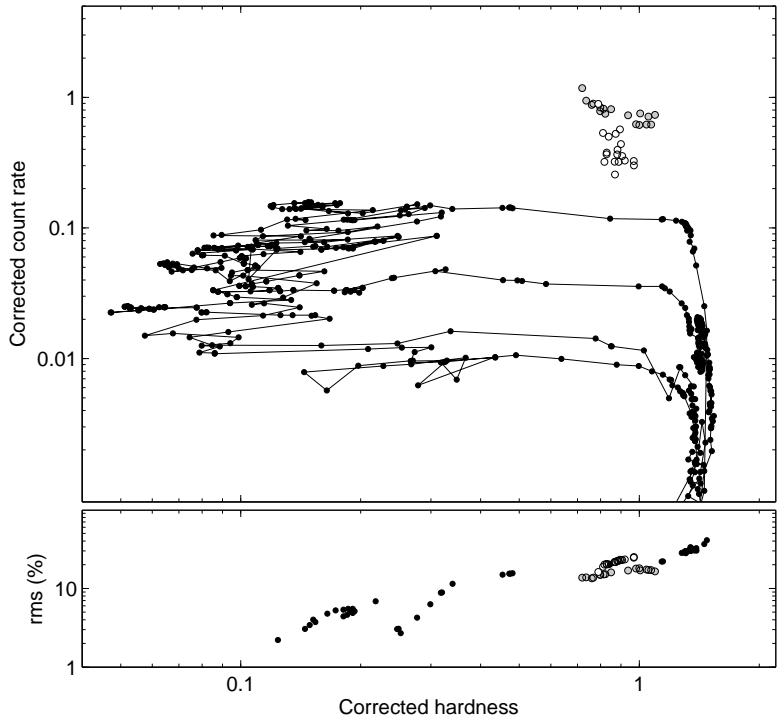}
\caption{
Top panel: HID of two ``plateau'' observations of GRS~1915+105, plotted over the HID of the 2002/3 and 2004/5 outbursts of GX 339--4 (black circles). The GRS~1915+105 points correspond to the observations of classes $\chi_1$ and $\chi_3$ (gray circles) and class $\chi_2$ (empty circles) from \cite{Belloni1915}. Note that the GX~339-4 points have been shifted to bring the source to the same distance as GRS~1915+105.
Bottom panel: HRD of the same observations, plotted over a few of the GX~339-4 from the 2002/3 outburst (see bottom panel of Fig. \ref{fig:hid339})
}
\label{fig:hid_1915}
\end{figure}

However, one can accumulate the intensity/hardness/rms of observations during the so-called ``plateaux''\index{plateaux}, which are long (one to few months) intervals of state C, usually followed by a major radio ejection (\cite{FenderBelloni}). During these observations, the flux only shows white noise on time scales longer than one second and the hardness can be compared with other sources as the thermal component is not dominant in the energy spectrum. Figure \ref{fig:hid_1915} shows the points of the observations corresponding to two of the plateaux examined in detail by \cite{Trudo}: the long one 1996 Oct--1997 Apr and the shorter during 1997 October.  The first one corresponds to variability class $\chi_2$, the second to classes $\chi_1$ and $\chi_3$ (see \cite{Belloni1915}). The HID shows points which lie, as expected, in the HIMS region, with the $\chi_1$ points at a higher count rate, probably as a result of a higher accretion rate. As in the case of Cyg~X-1, the count rate has been corrected also for the difference in distance between the two sources and the high brightness of GRS~1915+105 even during plateaux is evident.
Their distribution is elongated in count rate more than in hardness, at variance with the points of GX~339-4. In the HRD, the points from classes $\chi_1$ and $\chi_3$ overlap with those from GX~339-4, while $\chi_2$ points lie above. This was also noted by  \cite{Trudo}, who showed that the difference is due to the presence of an additional broad component at high-frequencies (50-100 Hz). This is the same component that was associated to lower-kHz QPO oscillations from neutron-star X-ray binaries (see \cite{bpk}). Moreover, the $\chi_3$ points correspond to a plateau with higher radio flux than that observed for $\chi_2$ \cite{FenderBelloni,Trudo}. GRS~1915+105 does therefore show some differences with other black-hole transients while in its HIMS, but overall the emerging picture is compatible.

\section{Neutron-star binaries}
\label{sec:7a}

\begin{figure}
\centering
\begin{tabular}{cc}
\includegraphics[height=4.5cm]{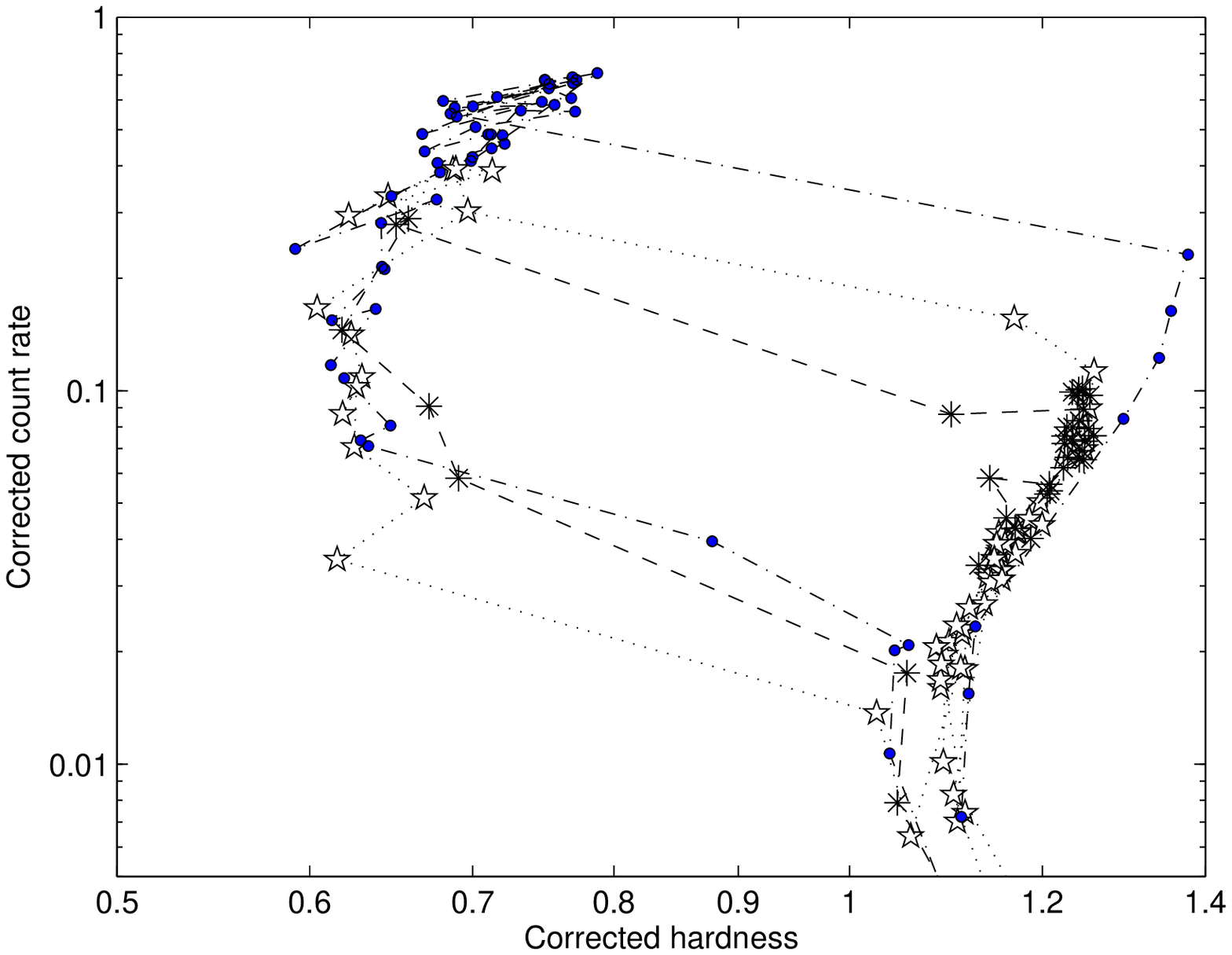} &
\includegraphics[height=4.5cm]{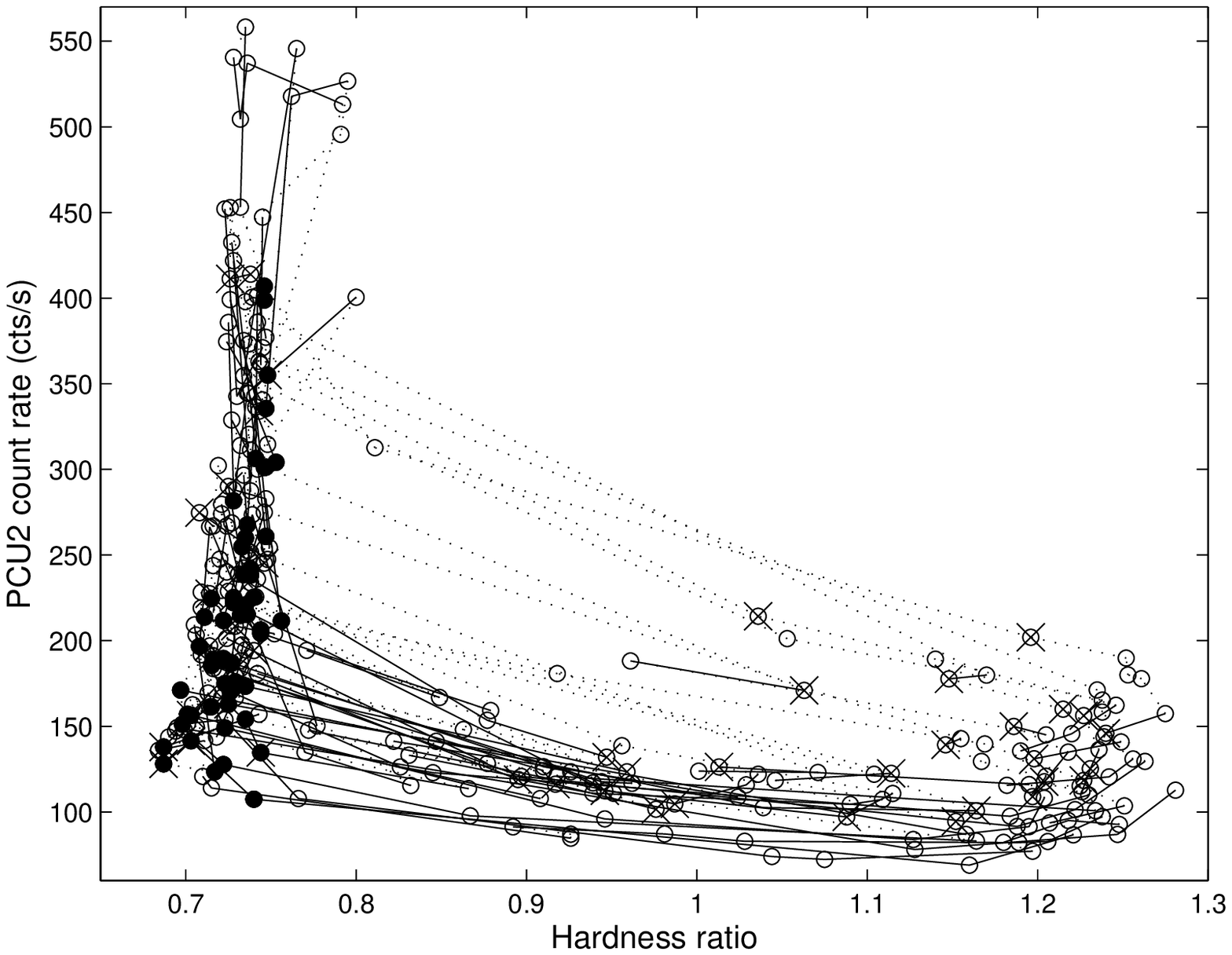}\\
\end{tabular}
\caption{
Left panel: HID for three outbursts of the transient neutron-star binary Aquila~X-1 as observed by RXTE/PCA. Stars, circles and asterisks indicate outbursts from 1999, 2000 and 2004 respectively. 
Right panel: HID from 305 RXTE/PCA pointings of the persistent neutron-star binary 4U~1636-53. The lines connect the observations in time sequence: solid line means time evolution from soft to hard, dotted line the reverse. From \cite{bmh2}.
}
\label{fig:ns}
\end{figure}

Many neutron-star low-mass X-ray binaries display a transient nature similar to that of black-hole systems (see \cite{vdkbook}). Their outbursts can also be characterized by the presence of two states, a hard and a soft one, with the hard state associated to low-flux intervals at the beginning and end of the outburst. Once a HID is produced, the similarity appears even stronger. Examples of HIDs for Aquila~X-1 can be found in \cite{vdkbook, koerdingscience} (see also Chaps. 4 and 5 of this book). A HID produced in the same way as the ones shown above can be seen in Fig. \ref{fig:ns} (left panel). It represents three well-sampled outbursts of Aql~X-1\index{Aql~X-1} as observed by RXTE. The similarity with Fig. \ref{fig:comparison_339}. The similarities between X-ray outbursts of black-hole and  neutron-star transients were explored by \cite{maccarone,yu1,yu2} and clearly suggest that the overall phenomenon of outburst evolution is similar. 
A similar evolution was recently found in the X-ray emission of a persistent neutron-star X-ray binary, 4U~1636-53\index{4U~1636-53} \cite{bmh2}. Systems of this class, called ``atoll sources'' alternate two states, characterized by spectral and timing features, a soft and a hard one, and accretion rate is thought to be higher in the soft state than in the high state \cite{vdkbook}. During the RXTE lifetime, 4U~1636-53 started long-term oscillations with a period of $\sim$45 days, which decreased with time to about $\sim$30 days \cite{shih,bmh2}. These oscillations correspond to regular state transitions: the production of a HID showed that the source travels a counter-clockwise path very similar to that of transients (see Fig. \ref{fig:ns}, right panel) \cite{bmh2}. The similarities were also extended to cataclysmic variables (see \cite{koerdingscience} and Chap. 5).

\section{Active Galactic Nuclei}
\label{sec:8}

The similarities between galactic X-ray binaries and AGN\index{AGN} are an important tool for the study of both systems. The inner region of the accretion disk is expected to be independent of the nature of the accreting system. When appropriate scaling laws are applied, comparison must be possible (see e.g. \cite{Fendermqw}).
The aperiodic timing properties of AGN have been studied in detail with long-term projects, needed because of the long time scales involved (see Chap. 8). From the analysis of twelve years of monitoring campaigns of AGN, two Seyfert 1 systems can be identified displaying variability properties similar to those of the HIMS. These are Ark~564\index{Ark~564} and Ton~S180\index{Ton~S180} \cite{Ian_ark564,Edelson}. 

A similar approach can be attempted through spectral/hardness analysis. However, there is a fundamental problem to be addressed before producing an AGN HID: the temperature of the thermal component from the inner radius of an optically thick accretion disk scales with the black-hole mass as $M^{-1/4}$. Therefore, for AGN this component does not appear in the X-ray band and the resulting HID would not be comparable to that of a galactic system. Recently, a comparison has been proposed based on `disk-fraction luminosity diagrams' (DFLDs), where in place of an X-ray color the fraction of the overall spectral distribution attributed to the accretion disk is used \cite{Elmar_hid}. Although this is a promising approach, it requires a very different type of analysis and is not directly comparable with BHB. 

\begin{figure}
\centering
\includegraphics[height=11cm]{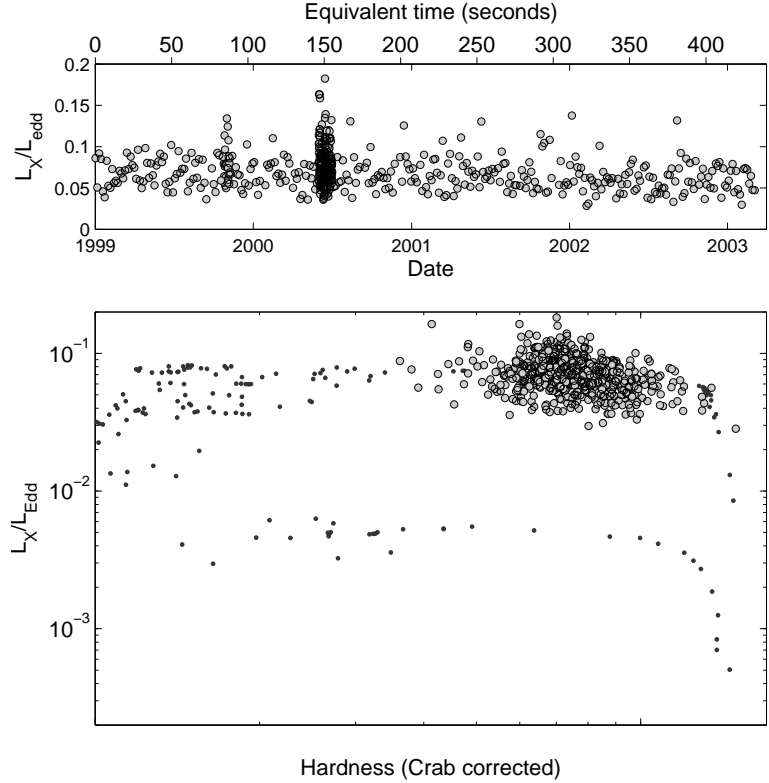}
\caption{
Top panel: RXTE/PCA light curve (one point per observation) of the Seyfert 1 AGN Ark~465. The Y-axis is in units of Eddington luminosity for a mass of 3.2$\times 10^6$M$_\odot$, with no bolometric correction from the 3.8--21.2 keV band. The top axis indicates the corresponding dynamical time scale if the mass of the black-hole were 10 M$_\odot$.
Bottom panel: corresponding HID, plotted over the 2002/3 points of GX~339-4.
}
\label{fig:hid_agn}
\end{figure}

The HID of a galactic system is made of hard observations, soft observations, and transitions between them. The softening at the beginning of a transient outburst is due to two combined effects: the appearance of a strong thermal disk component and the steepening of the hard component. The hardening at the end of the outburst is the reverse effect. This means that even ignoring the disk component, the HID would have a similar shape, although the source would never reach very soft values of hardness. In other words, after the thermal disk is removed, the LHS points would not change (no strong disk component is detected there) and the HSS points would become considerably harder (and weaker, since here the flux is almost all thermal disk). In this respect, it appears meaningful to produce HIDs for AGN using the same procedure used for binaries. From the analysis of all Seyfert~1 AGN in the RXTE database, two systems emerge as rather different in their HID: not surprisingly, they are Ark~564 and Ton~S180, the same ones singled out by the timing analysis. As an example, Fig. \ref{fig:hid_agn} shows the results for Ark~564. The top panel shows the light curve (one point per RXTE observation) over the period 1999-2003. The flux here is in the 3.8--21.2 keV, renormalized to the flux at Eddington accretion rate measured in the same band (i.e. an Eddington-normalized flux without a bolometric correction). The top axis shows the same $\sim$4-year scale linearly scaled from the mass of the black hole in Ark~564 to 10 $M_\odot$: four years for the AGN correspond to 500 seconds for a typical galactic black-hole. The bottom panel is the HID with the same flux as in the top panel and a hardness corrected for the spectrum of the Crab nebula at the time of the observation, in oder to compensate for gain changes throughout the RXTE lifetime. The dots show the points for the 2002/3 outburst of GX~339-4, where the hardness is similarly corrected. It is evident that the points of Ark~564 are distributed along the horizontal HIMS line of the galactic binary. As explained above, their range is naturally limited and cannot reach very low values of hardness as those of the HSS of GX~339--4. 

Notice however that GX~339-4 moved through the HID branch on a time scale of days and in a right-to-left direction. Ark~564 moves erratically and its points span a mere 500 seconds after correction for the black-hole mass. This means that, while the average value of hardness indicates Ark~564 as a good candidate for an AGN in the HIMS, the elongated shape of the distribution represents change on a different time scale than what observed from GX~339-4. In this sense, AGN offer a unique chance to explore short time scales for which galactic sources do not have sufficient statistics.

While the two `intermediate' AGN can be identified both from timing and hardness analysis, there is a problem for the others. All Seyfert~1 AGN are located on the hard branch, corresponding to the LHS, while time variability suggests that they are all in the HSS (see e.g. \cite{Ian_scaled}).

Recently, a 1-hour QPO was discovered in the Narrow-Line Seyfert 1 RE~J1034+396\index{RE~J1034+396} through a long XMM observation (see Chap. 8), opening an important window for comparison of timing properties of supermassive and stellar-mass systems \cite{agnqpo}. Given its relatively high frequency, this oscillation would correspond to high-frequency QPOs in BHBs.

\section{Models and interpretation}
\label{sec:9}

In the previous sections, I have presented a phenomenological picture for the evolution of hardness and timing properties in galactic black-hole transients and have compared it with those of other stellar-mass systems and AGN. In order to derive a physical picture, energy spectra must be extracted and theoretical models applied. It appears as the timing properties are crucial, but to date we do not have a complete theoretical framework for them (see \cite{vdkbook} for a review). 
The emission mechanisms associated to the LHS and the HSS were discussed in Chap. 2, together with the geometry of the accretion flow in those states. Here, I will once again concentrate on the variations along the diagram, with particular emphasis on the state transitions. The picture presented in the previous sections is based on the emission in the 4-20 keV energy band, while a physical interpretation cannot be attempted without information on a much broader energy range. 

\subsection{Energy spectra}

All outbursts probably start with a LHS period, although in some case we do not observe it because it is too short. The general direction of accretion rate along the path in Fig. \ref{fig:states} is vertical: higher points correspond to a higher accretion rate. When a source moves horizontally, accretion rate should be higher on the left, but this is clear only along the main transitions at the top and bottom of the HID.
Since accretion rate drives the movement along in the HID, a different physical parameter must be responsible for the position, i.e. accretion rate threshold, of the LHS-HIMS transition, which can vary between outbursts of the same source. The nature of this second parameter is very important, but still unclear (see Chap. 9).

Over the right branch in the HID, the broad-band energy spectrum is hard, with a high-energy cutoff around 100 keV see \cite{Grove}. However, an INTEGRAL\index{INTEGRAL} observation of GRO~J1655-40 has failed detecting a high-energy cutoff \cite{Caballero}. In the few cases when the level of interstellar absorption is low, a soft thermal disk component is also detected, with a large disk inner radius \cite{Filippo1118,Mc1118}. Recently, a hotter thermal disk has been detected from a few sources, suggesting that the inner disk radius might always be close to the innermost stable orbit \cite{Miller1753,Miller339}. 
The issue of the truncation of the inner disk radius is still unsolved. A Suzaku observation of Cygnus~X-1 in the LHS from 2005 shows that the inner disk radius is indeed large, larger than that measured in the HSS \cite{Suzakucigno}. Moreover, a re-analysis of the RXTE and XMM data for Swift~1753.5-0127 showed that a receded disk cannot be ruled out \cite{hiemstra}.
The hard component is usually interpreted as the result of thermal Comptonization or of a combination of thermal/non-thermal Comptonization (e.g. \cite{Ibragimov}). 
Over the left branch, in the HSS, the energy spectrum is dominated by a hot thermal accretion disk, together with a (variable) hard component. This hard component has been observed as a power law without a cutoff up to 1 MeV \cite{Grove}. 

What is the shape of the broad-band energy spectrum in the HIMS/SIMS and does anything change during the fast transitions? In other words: how does the energy spectrum change from the LSS to the HSS? The energy spectrum in these states is a combination of those of LSS and HSS. The thermal disk is clearly present and it is responsible for part of the horizontal excursion in the HID.
Whether the inner disk radius moves inward as a source softens through the HIMS is still debated (see above). Significant radius changes were observed from GRS~1915+105 \cite{Belloni1915}.
 The hard component steepens as the disk fraction increases. Across transitions, the 2-20 keV spectrum does not change appreciably. For transients, a broader-band spectrum of good quality is only possible for the hard-to-soft transitions at high luminosities. 
 
 Recent RXTE/INTEGRAL observations of GX~339-4\index{GX~339-4} have given conflicting results as to whether the high-energy cutoff increases/decreases/disappears across a HIMS--SIMS transition \cite{Belloni_INTEGRAL,delsanto2009}. Recent observations of GX~339-4 \cite{motta2009} and GRO~J1655-40 \cite{joinet2008} show that the situation is more complex than a simple increase or disappearance of the cutoff. More observations are needed. These observations are difficult as the transition is fast and can only be predicted with an accuracy of weeks. More information is available from GRS~1915+105, which as we saw above is only found in intermediate states. Simultaneous RXTE/OSSE\index{OSSE} observations showed spectra without a measurable cutoff up to $\sim$600 keV, which could be interpreted with a hybrid thermal/non-thermal Componization model \cite{Zdz1915}. An RXTE/INTEGRAL campaign on GRS~1915+105 showed that the energy dependence of the type-C QPO (a HIMS signature) can be explained also with a hybrid emission model \cite{Jerome1915}.
In addition, quasi-simultaneous ASCA/RXTE/OSSE data of XTE~J1550-564 during the HIMS were found also to require a hybrid thermal/non-thermal model \cite{GierDone1550}.
The RXTE database of GX~339-4, including the 2002/3 outburst shown in the previous sections, has been analyzed with spectral models by \cite{Zdz339}.

The HIMS-SIMS transition corresponds to the crossing of the ``jet line'', which approximately marks the time of the ejection of a fast relativistic jet from the system (see the following chapters). This correspondence is not precise, as shown by \cite{FenderHomanBelloni09}, see Chap. 5. However, 
this event takes place in a short interval of time (minutes to hours). It is possible that the transient annihilation line detected from GS~1124-68\index{GS~1124-68}, which was seen only over an interval of $\sim$12 hours, was indeed associated to such a transition, since the observation took place well within the Very-High State of the source, corresponding to HIMS/SIMS \cite{Goldwurm}. However, no evidence for such a spectral feature was found in the INTEGRAL data of the 2004 transition of GX~339-4.

As to the anomalous state, detailed spectral fitting needs to be performed. Here the flux is high, associated to high luminosity and, since the energy spectrum is soft, mass accretion rate is probably also high. GRO~J1655-40 showed a well populated anomalous branch in both its outburst covered by RXTE. XTE~J1550-564 in 1998 showed a very high flux in a single observation, exceeding a flux of 6 Crab \cite{Sobczak1550}, which could be interpreted as belonging to the same ``state'' \cite{fbg04}. What these observations have in common are a high inner-disk temperature, a steep hard component, a very small derived inner disk radius and a small percentage contribution of the disk to the overall flux \cite{Sobczak1655,Sobczak1550}.
The same effect was measured in 4U 1630-47 in a number of RXTE observations \cite{Tomsick1630}. These were interpreted as a signature to the transition to a slim disk regime \cite{Tomsick1630,Watarai}. The issue of the shape of the soft component has been discussed by various authors; in particular, the distortion effects due to Comptonization have been investigated \cite{DoneKubota2006}. A comprehensive physical picture of the accretion flow onto black holes and neutron stars has been proposed \cite{Done2007}, but it is not yet clear how this corresponds to the states presented above and the phenomenology is mixed with the modeling.

\subsection{Time variability}

The fast aperiodic variability appears to be a crucial element for our understanding of the accretion (and jet ejection) properties in black-hole binaries. In particular, the most important state-transition (HIMS--SIMS) associated to the crossing of the jet line is identified only through the changes in the timing properties. 

The PDS of the LSS and the HIMS is compatible with a smooth evolution of power-spectral components. The band-limited noise components and the type-C QPO follow very tight correlations which can be extended to neutron-star binaries and possibly to cataclysmic variables \cite{wvdk,bpk,Warner,vdkbook}. It is clear now that for these features, the distinction between noise and QPO components is somewhat arbitrary (see \cite{bpk}). An interesting theoretical approach that does not discriminate between them has been presented, based on a dynamical model that investigates the filtering effect of the presence of a transition radius in the accretion disk \cite{PsaltisNorman}.
Moreover, the frequencies of these component correlate with spectral parameters. A strong general correlation was found between the frequency of the type-C QPO and the photon index of the hard power-law component in the energy spectrum \cite{Vignarca}. This correlation has been suggested as a possible way to estimate the compact-object mass from X-ray data (see e.g. \cite{LevMass}). The fact that these correlations appear to extend also to the HSS indicates that the nature of the fundamental frequencies driving the variability is the same from the LHS across the HIMS to the HSS. It is also possible that the actual emission process is the same in the hard and soft states. This would indicate that the ``corona'' which some models associate to the hard-state emission would be present in the soft state. More analysis is needed in order to establish this fact.

Definitely different is the case of the SIMS. Here the timing properties are different: there is no band-limited noise and the QPOs have different characteristics, notably are rather stable in frequency. At the same time, high-frequency QPOs are at times observed. What happens during this state, the onset of which is associated to the launch of fast relativistic jets, is still unexplored. The sharp and fast transitions observed from and to this state show that probably some type of instability is at work here, but no modeling has been attempted.

Unlike in the HID, the source track in the HRD does not show evidence of hysteresis (see Figs. \ref{fig:hid339} and \ref{fig:states}). The main difference between the high-flux and low-flux branches is related to the small extent of the SIMS in the latter. However, detailed analysis does show that the two HRD paths are different \cite{Belloni05}. Therefore, the timing properties seem to depend only weakly on the flux, providing that the energy spectrum is the same. This contributes to indicate that time variability provides a powerful and direct approach to basic parameters in the accretion flow.
\footnote{Notice that the different energy dependence of the fractional integrated variability corresponding to the different states \cite{Gzdz05} means that the HRD looks different at different energies. In particular, the different energy dependence in the LHS and HIMS breaks the continuity of the path in the HRD at high energies.}
This also applies to the elusive HFQPOs, which are observed in or near the SIMS.  Unfortunately, the scarcity of detections does not allow more thorough investigations on the nature of these oscillations and their relation to the spectral characteristics of the associated emission.

\section{Conclusions: how many states?\index{states}}

The phenomenology of the evolution of outbursts of black-hole transients appears complex, but not impossible to treat. The use of diagrams such as HID and HRD allows to disentangle most of the properties and compare the behavior of different systems. This paradigm can also be applied to other stellar-mass systems and even to supermassive objects.
Although the state classification presented in the previous sections manages to capture all the essential timing/hardness properties of these systems, a major question which still remains open is that of physical states. The original picture, still widely applied today, interprets observations in terms of two separate physical states: a hard one dominated by thermal Comptonization and a soft one dominated by an optically thick accretion disk with a contribution by a non-thermal hard component. The study of state-transitions and global timing properties suggests that the differences between hard and soft states are not so marked. They could be considered as evolving smoothly from the hardest energy spectra associated to strong variability to the softest energy spectra with only a few \% variability. The energy spectra also do not seem to show strong transitions: the INTEGRAL observation of the HIMS-SIMS transition of 2004 shows a hint of a change in the high-energy cutoff, but the spectral index at low energies is unchanged (see also \cite{Nespoli03)}).
The only interruption is the presence of the SIMS, which is marked by a fast drop in variability and the appearance of specific types of QPOs. This picture suggests that the only physically different state is the SIMS, which is also associated to the ejection of relativistic jets. As mentioned at the beginning, this picture and the corresponding state classification are meant to only to present a coherent picture: the development and application of theoretical models will identify physically meaningful states and lead to a full classification. This will be possible only if the complete evolution of transient black-hole binaries is considered.

\subsection*{Acknowledgements}

A large number of colleagues have contributed to the shaping of the concepts which I have put in this chapter, and of course also to the results. Impossible to list them here, but easy to identify them as co-authors of many papers (and of other chapters of this book). The main one is certainly Jeroen Homan, whose work has started the new wave in black-hole states and is now consolidating it.

%
%

%
%

\end{document}